# Multinary Alloying Suppresses Defect Formation in Emerging Inorganic Solar Cells

Jiangjian Shi[1†], Jinlin Wang[1,3†], Fanqi Meng[4†], Jiazheng Zhou[1,3†], Xiao Xu[1,3], Kang Yin[1,3], Licheng Lou[1,3], Menghan Jiao[1,3], Bowen Zhang[1,3], Huijue Wu[1], Yanhong Luo[1,3], Dongmei Li[1,3], Qingbo Meng[1,2]*

[1] Key Laboratory for Renewable Energy, Beijing Key Laboratory for New Energy Materials and Devices, Institute of Physics, Chinese Academy of Sciences, Beijing 100190, P. R. China.

[2] Center of Materials Science and Optoelectronics Engineering, University of Chinese Academy of Sciences, Beijing 100049, China.

[3] School of Physics Science, University of Chinese Academy of Sciences, Beijing 100049, P. R. China.

[4] State Key Lab of New Ceramics and Fine Processing, School of Materials Science and Engineering, Tsinghua University, Beijing, 100084 China.

†These authors contributed equally to this work.

*Corresponding author. Email: qbmeng@iphy.ac.cn

**Abstract:** The Cu$_2$ZnSn(S, Se)$_4$ (CZTSSe) emerging inorganic solar cell is highly promising for accelerating the large-scale and low-cost applications of thin-film photovoltaics. It possesses distinct advantages such as abundant and non-toxic constituent elements, high material stability, and excellent compatibility with industrial processes. However, CZTSSe solar cells still face challenges related to complex defects and charge losses. To overcome these limitations and improve the efficiency of CZTSSe solar cells, it is crucial to experimentally identify and mitigate deep defects. In this study, we reveal that the dominant deep defect in CZTSSe materials exhibits donor characteristics. We propose that incomplete cation exchange during the multi-step crystallization reactions of CZTSSe is the kinetics mechanism responsible for the defect formation. To address this issue, we introduce an elemental synergistic alloying approach aimed at weakening the metal-chalcogen bond strength and the stability of intermediate phases. This alloying strategy has facilitated the kinetics of cation exchange, leading to a significant reduction in charge losses within the CZTSSe absorber. As a result, we have achieved a cell efficiency of over 14.5%. These results represent a significant advancement for emerging inorganic solar cells and additionally bring more opportunities for the precise identification and regulation of defects in a wider range of multinary inorganic compounds.

The ongoing exploration of diversified photovoltaic materials has created more opportunities for the efficient, low-cost, and environmentally-friendly utilization of solar energy.[1-2] The emerging multinary chalcogenide, $Cu_2ZnSn(S, Se)_4$ (CZTSSe), is being developed as a promising candidate to promote the large-scale and low-cost application of thin-film photovoltaics, by leveraging its advantages of high constituent element abundance, nontoxicity, outstanding material stability, and excellent industrial compatibility.[3-7] However, the multinary element components in CZTSSe have also led to complex atomic self-doping and intrinsic defects, resulting in severe non-radiative charge recombination and significant photoelectric conversion efficiency (PCE) loss in solar cells.[2,8,9]

Identifying and regulating intrinsic defects is currently the crucial requirement for realizing high-performance CZTSSe solar cells.[1,2] The low formation energy of self-doping is generally considered as the physics mechanism of defect formation.[2,8,9] To suppress defects, a variety of thermodynamics approaches, such as element content fine optimization, Ag/Cu substitution doping, and Fermi energy level modification, have been successively employed to increase the defect formation energy.[10-13] However, these efforts have not been able to help CZTSSe solar cells overcome the impasse of stagnant efficiency. In the past ten years, the cell record PCE has only improved by 0.4%, from 12.6% to 13.0%. This may be due to several possible reasons. Firstly, the killer defect in CZTSSe has yet been determined experimentally, which makes the defect regulation lack clear pertinence. Secondly, the widely used thermodynamics equilibrium mechanism cannot completely reflect the defect formation in CZTSSe because its crystallization is dominated by solid-state heterogeneous reactions.[14-15] Additionally, the demands on phase structure and carrier doping of CZTSSe have also restricted the regulation of the defect formation energy.[2,9]

Herein, we present a kinetics-based approach for suppressing the formation of deep defects in CZTSSe, resulting in a remarkable improvement of the cell PCE up to 14.6%. By utilizing a cell parameter analysis method, we identify the donor-type defect, most possibly the Sn/Zn antisite ($Sn_{Zn}$), as the primary deep defect in CZTSSe. We propose that the incomplete cation exchange in the multi-step crystallization reactions of CZTSSe is the primary kinetics mechanism responsible for the formation of this defect. These findings lead us to introduce elemental alloying to reduce the material stability and metal-chalcogen bond strength of the intermediate phases, thereby facilitating cation exchange. This approach has significantly suppressed the deep defect-induced charge loss in the CZTSSe absorber and resulted in an impressive step forward in the cell PCE. The presented defect analysis method and understanding of the defect formation mechanism can also guide defect regulation in a wider range of multinary inorganic compounds.

**Defect identification in CZTSSe**

Identifying the type of killer defect and its formation mechanism is the prerequisite for precise control of semiconductor defects. Here, we propose using the correlations of cell performance parameters with the defect to evaluate the type of defect in CZTSSe. This approach overcomes the inconsistency of defect measurement in previous works.[14] Using device simulation (Fig. S1-S3 and Table S1),[16-17] we found that cells with different defect types (acceptor or donor) are clearly segregated into two regions in the $J_{SC}$/$V_{OC}$/FF-PCE characteristics graph ($J_{SC}$: short-circuit current density, $V_{OC}$: open-circuit voltage, FF: fill factor), as shown in Fig. 1a. Specifically, for the cells with donor defects, the $J_{SC}$/$V_{OC}$/FF increases much more obviously when the PCE decreases, mainly due to the charge compensation between donor defects and majority holes. These behaviors make the $J_{SC}$/$V_{OC}$/FF-PCE characteristics a reliable criterion for identifying the deep defect type.

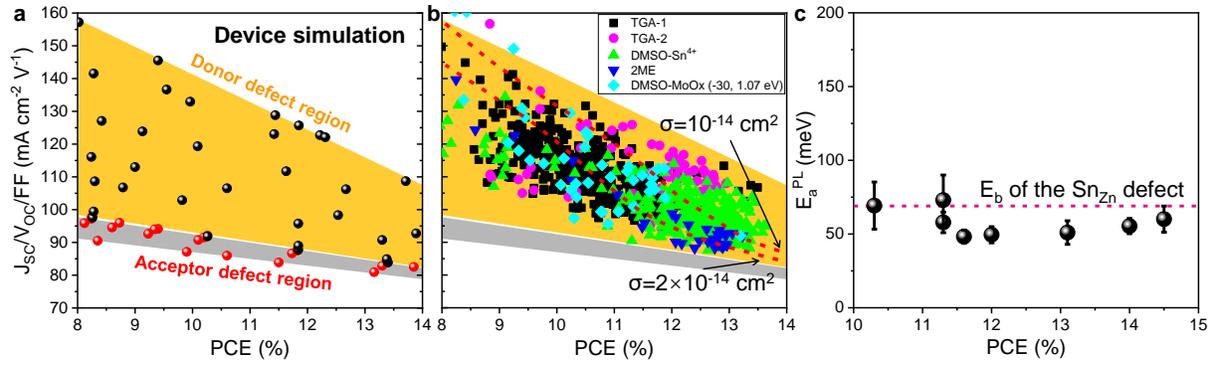

**Fig. 1. Defect type analysis of CZTSSe.** (a) Impact of defect type (donor and acceptor) on the cell $J_{SC}$/$V_{OC}$/FF-PCE characteristics. (b) A comparison of experimental data to the simulation results. The experimental data were collected from ~1000 cells fabricated with different methods in our lab (TGA-1: thioglycolic acid-water solution method, DMSO-$Sn^{4+}$: dimethyl sulfoxide solution with $SnCl_4$ as the Sn source, 2ME: 2-Methoxyethanol solution method, DMSO-$MoO_x$: back interface modification by $MoO_x$ and $SnCl_2$ as the Sn source). (c) Photoluminescence (PL) quenching activation energy ($E_a^{PL}$) of the fabricated CZTSSe films wirth different cell PCEs. The calculated carrier-capture energy barrier ($E_b$) of the $Sn_{Zn}$ defect is given for reference. Error bar: fitting error.

We subsequently collected performance parameters on approximately 1000 cells from our laboratory and compared with the simulation results.[18,19] We assume that the fluctuation in the PCE of these cells is mainly caused by variations in deep defects in CZTSSe, as the formation of defects is the most uncontrollable variable in experiments. Figure 1b shows that nearly all experimental data fall within the donor-defect region, indicating that the cell's PCE is mainly determined by the deep donor defect. This finding differs from previously proposed scenarios of acceptor-type deep defects[10,20] but is supported by recent theoretical predictions.[9] Additionally, the similar distribution trend of $J_{SC}$/$V_{OC}$/FF-PCE data between different cell groups suggests that these CZTSSe absorbers have the same defect characteristics despite being fabricated by different methods and operators. The defect carrier capture cross-section ($\sigma$) of these CZTSSe absorbers is estimated to be in the order of $10^{-14}$ $cm^2$. We also measured the photoluminescence

(PL) quenching activation energy ($E_a^{PL}$) of the CZTSSe films, which gave values ranging between 50 and 70 meV (Fig. 1c). These small $E_a^{PL}$ values agree with the large σ and are very close to the predicted carrier-capture energy barrier of $Sn_{Zn}$ donor defect.[8,9,21] Therefore, we determine that the primary deep defect in these cells is $Sn_{Zn}$.

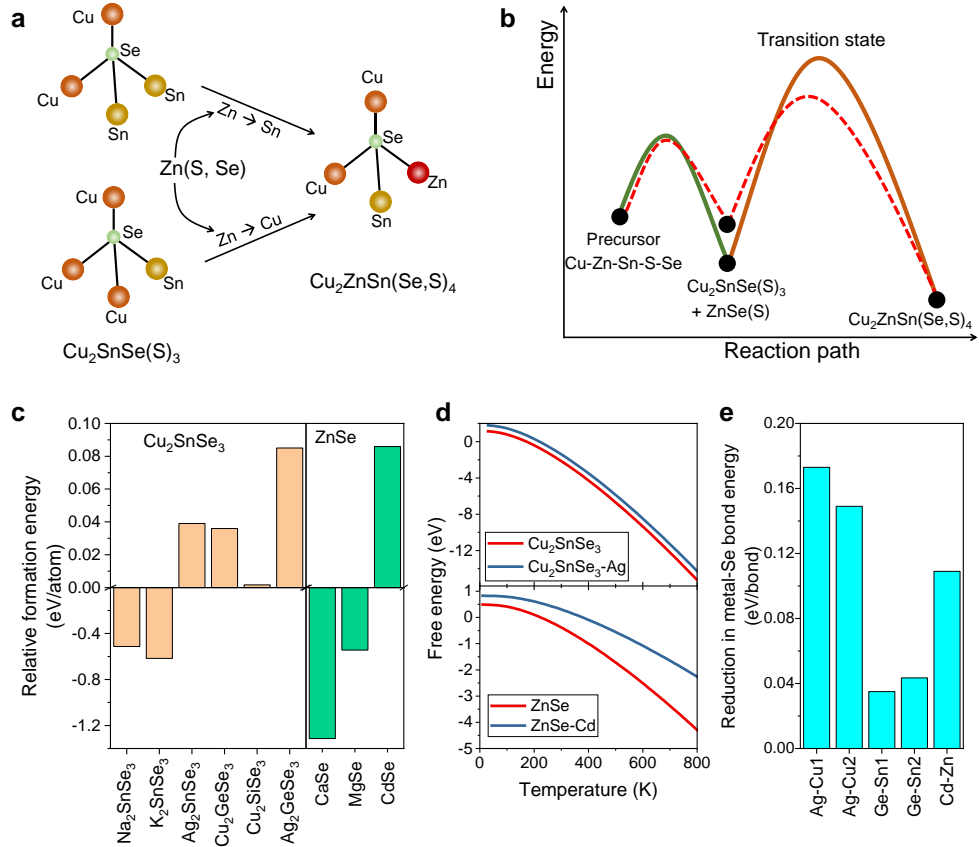

**Fig. 2. Kinetics mechanism and regulation of $Sn_{Zn}$ defect formation.** (a) Schematic diagrams of cation exchange processes in the phase evolution from CTSSe to CZTSSe and (b) of the multi-step reaction pathways. (c) Relative formation energy of $A_2BSe_3$ (A: Na, K, Ag, Cu; B: Si, Ge) and MSe (M: Ca, Mg, Cd) compounds compared to $Cu_2SnSe_3$ and ZnSe, respectively. (d) Gibbs free energy of $Cu_2SnSe_3$, ZnSe and their alloyed compounds. (e) Reduction in the metal-Se bond energy of $Cu_2SnSe_3$ and ZnSe when they are partially alloyed.

It was previously predicted that the cell PCE limited by the $Sn_{Zn}$ defect under thermodynamic equilibrium can reach 20% or higher,[8,9,22] which is significantly better than currently achieved results in experiment. This implies that thermodynamics equilibrium does not determine the formation of defects in current cells. In our opinion, the incomplete cation exchange during the CZTSSe multi-step solid-phase crystallization reactions is a more important factor in the formation of $Sn_{Zn}$ defect formation. During the initial crystallization stage of CZTSSe, trinary $Cu_2Sn(S, Se)_3$ (CTSSe) and binary Zn(S, Se) intermediate phases are usually formed earlier at lower temperatures and Se concentrations.[23-26] To form the final CZTSSe, Zn needs to release from Zn(S, Se) and partially substitute Cu and Sn in CTSSe, as schematically shown in Fig. 2a-b. Although this reaction is thermodynamically favorable at high temperatures, the energy barriers of metal-Se bond breaking and atom insertion in these processes still limit the kinetics of Zn/Sn and Zn/Cu exchanges, thus leading to the formation of $Sn_{Zn}$ defect in the final CZTSSe. To reduce the occurrence of this defect, cation exchange needs to be facilitated, by decreasing intermediate-phase stability and reducing the cation-exchange barrier, as shown by the dashed line in Fig. 2b.

**Elemental alloying**

To achieve this goal, elemental alloying offers a promising solution as it can effectively modify the atomic bonding state of a material.[27-28] Initially, we used material formation energy ($E_{form}$) as a criterion to search suitable alloying elements. To avoid introducing carrier doping, only +1, +2 and +4 valent elements were considered. We collected and compared the $E_{form}$ of a variety of $A_2BSe_3$ (A: Na, K, Ag, Cu; B: Si, Ge) and MSe (M: Ca, Mg, Cd) compounds with that of $Cu_2SnSe_3$ and ZnSe (Table S2).[29] Fig. 2c shows that Ag or Ge-containing $A_2BSe_3$ compounds have a smaller $|E_{form}|$, making them potential alloying elements. CdSe in MSe also exhibits a similar trend. We further calculate the Gibbs free energy of CTSe, ZnSe and their alloyed compounds with these three elements. As shown in Fig. 2d, Ag alloying in CTSe increases the free energy by approximately 0.9 eV at 800 K, and Cd alloying in ZnSe increases the free energy by about 2.0 eV at 800 K. Ge has little influence on the free energy. We also calculated the change in the metal-Se bond energy of these compounds (Fig. S4-S6). The results shown in Fig. 2e indicate that the average energy of the metal-Se bond around the alloying site is reduced by approximately 0.16 (Ag), 0.04 (Ge) and 0.11 eV (Cd), respectively. For ZnSe, the 0.11 eV reduction accounts for one-fourth of its initial bond energy.[30] Hence, in the high-temperature reaction process, Zn can release from ZnSe and insert into the CTSe phase much more easily, facilitating the elimination of $Sn_{Zn}$ defect in the final CZTSSe.

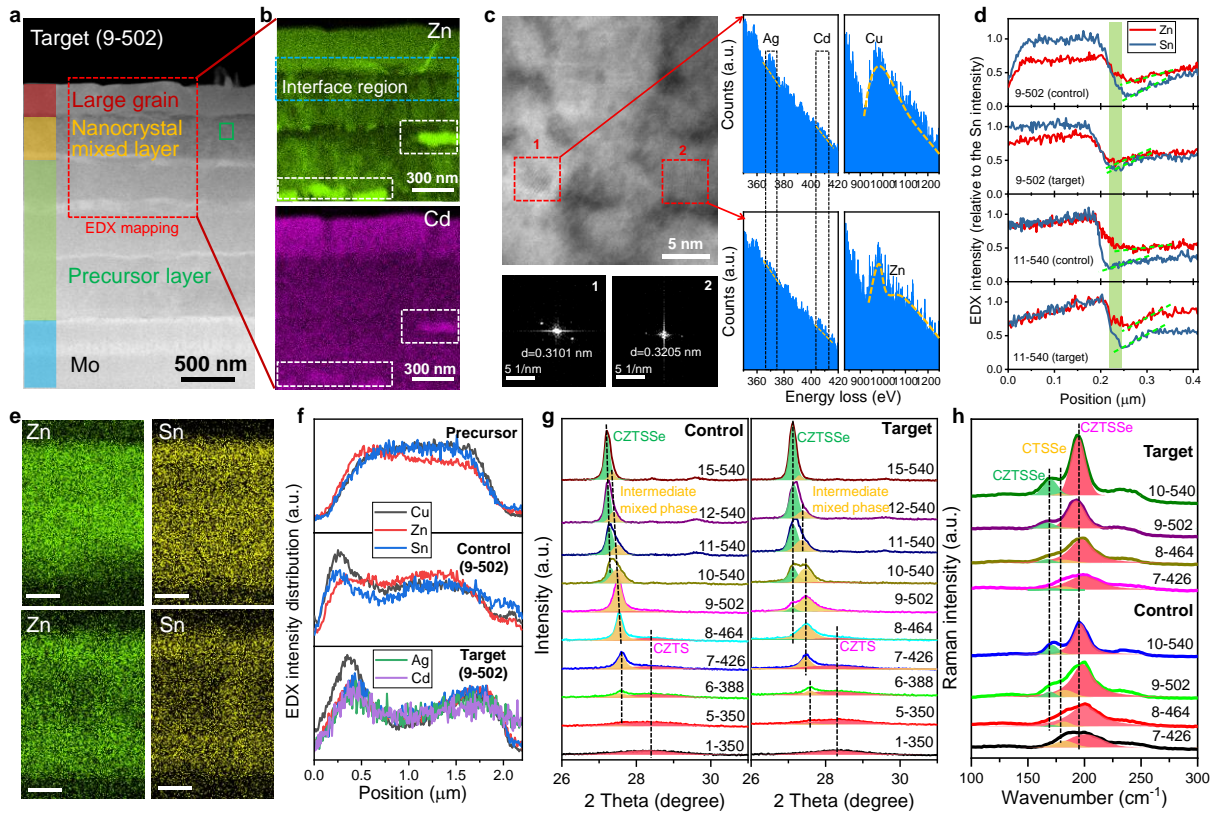

**FIG 3. Influence of elemental alloying on the solid-phase reaction in CZTSSe films.** (a) Cross-sectional HAADF STEM image of the target 9-502 film. (b) STEM EDX (Zn and Cd) images of a selected region of the target film. (c) HAADF image of a selected region in the nanocrystal mixed layer, and fast Fourier transform (FFT) patterns and electron energy loss spectra (EELS) of two observed nanocrystals. (d) Zn and Sn element distribution crossing the CZTSSe/mixed layer interface. The gradient of element intensity is depicted by dashed green

lines. (e) Cross-section SEM EDX mapping image and (f) profiles of the 9-502 films. Scale bar: 500 nm. Evolutions of X-ray diffraction patterns (g) and Raman spectra (b) of the films in the selenization process.

In the experimental study, the three aforementioned alloying elements were introduced into CZTSSe films via precursor solutions. The films were sampled at different stages of the selenization crystallization process (Fig. S7-S8) and were labeled according to the time-temperature point of sampling for clarity. The films were then characterized using a spherical aberration corrected scanning transmission electron microscope (STEM). Cross-sectional high-angle annular dark-field (HAADF) images (Fig. 3a and S9) reveal that the alloyed (target) and unalloyed (control) samples exhibit similar top-down crystallization modes and have the same crystal growth orientation along the [201] direction of the CZTSSe lattice (Fig. S9-S10). The top region of the films has transformed into large grains, while the lower region remains in a nanocrystal-amorphous mixed state (Fig. S11). In the energy dispersive X-ray spectroscopy (EDX) mapping image of the target 9-502 sample (Fig. 3b), the simultaneous enrichment of Zn and Cd elements is observed in some nanoscale regions. Electron energy loss spectroscopy (EELS) analysis reveals the coexistence of Ag/Cu or Cd/Zn in different nanocrystals within the mixed layer (Fig. 3c). Furthermore, the occupation of Ag at the Cu site in the top CZTSSe lattice is also confirmed (Fig. S12). These findings indicate that elemental alloying did occur in the initial crystallization reaction stage.

After alloying, it is apparent from Fig. 3d that the gradient of Zn element distribution at the interface between the top grain and the nanocrystal mixed layers has increased significantly, becoming similar to that of Sn. According to Fick's first law, this indicates that the velocity of Zn element insertion into the top crystals has been enhanced. Consequently, Zn acquires a more synchronized distribution with Cu and Sn elements throughout the film (Fig. 3e-f, S13-S14). In comparison, in the control sample, Zn exhibits low reaction activity with Cu and Sn and thus has been repelled to the middle region.

The X-ray diffraction (XRD) patterns evolution also indicates that the final-like CZTSSe phase appeared earlier in the target sample, and its XRD peak position remained constant during the entire selenization process (Fig. 3g, Fig. S15-S16). Additionally, the Raman spectra of the target sample shown in Fig. 3h display a more pronounced Zn-related lattice vibration (150-175 cm$^{-1}$) and a much weaker CTSSe signature (180 cm$^{-1}$).[22] The X-ray photoelectron valence-band spectra also demonstrate an enhancement of Zn-Se(S) hybridization in CZTSSe (Fig. S17).[31,32] These results suggest that elemental alloying has facilitated the exchange of Zn/Sn and Zn/Cu, leading to an accelerated formation of ordered CZTSSe phase. This, in turn, is expected to help suppress the formation of $Sn_{Zn}$ deep defect.

**Solar cell characterization**

We subsequently fabricated CZTSSe solar cells and analyzed their performance. The statistics of a batch of cells are presented in Fig. 4a and S18. The $J_{SC}$/$V_{OC}$/FF-PCE analysis revealed that all cells followed the donor defect characteristics. The target cell had an average PCE of about 13.5%, while the control cell had an average PCE of only about 10.5%. The PCE improvement was mainly attributed to the increase in $V_{OC}$ and FF, due to reduced charge recombination loss. We also fabricated cells with more alloying groups, as shown in Fig. S19. The best result was obtained when Ag, Ge, and Cd were synergistically used (Fig. S20), where Ag/Ge modified CTSSe and Cd modified Zn(S, Se) phase, respectively. We also attempted to use Ag/Ge/Mg to alloy the CZTSSe, but obtained obviously lower PCE (Fig. S21). Thus, using the Eform to theoretically search for alloying elements is feasible. Furthermore, we attempted a post-deposition treatment (PDT) method to introduce these alloying elements, but it did not work well.

This implies that compared to final-state doping, it is more important for these elements to participate in and regulate the crystallization reactions.

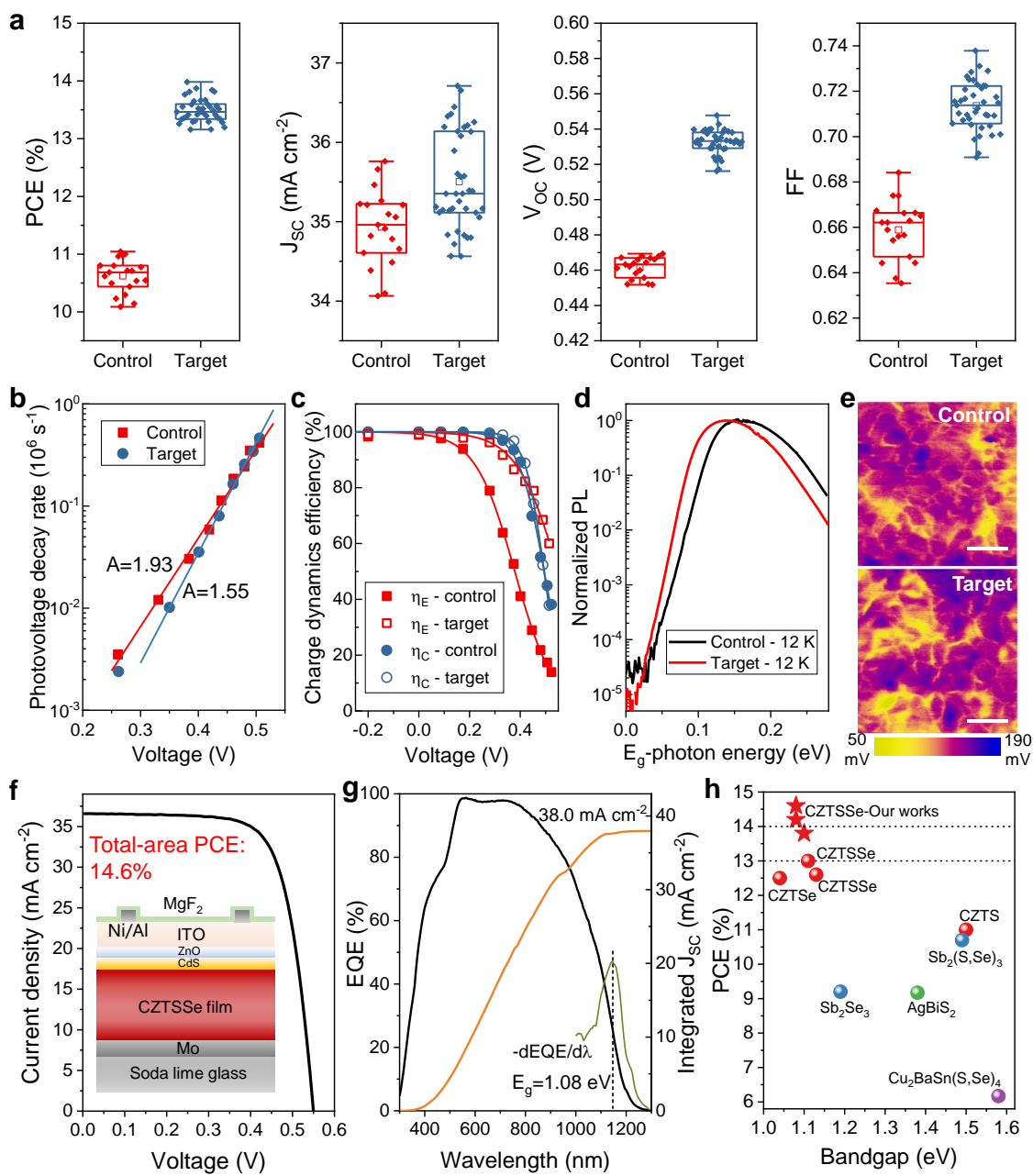

**Fig. 4. Solar cell characterization.** (a) Statistics of the cell performance parameters. (b) Voltage-dependent photovoltage decay rate of the cells. The solid lines depict the exponential fitting for determining the ideality factor (A). (c) Charge extraction ($\eta_E$) and collection ($\eta_C$) efficiencies of the cells with or without the elemental alloying. (d) Photoluminescence (PL) spectra and (d) temperature-dependent PL intensity of the CZTSSe films. (e) Surface contacting

potential difference (CPD) images of the CZTSSe films. Scale bar: 2 μm. (f) Current-voltage characteristics of the champion cell which has a total-area PCE of 14.6%. Inset: cell configuration. (g) External quantum efficiency (EQE) spectrum of the cell. (h) Champion performance of different emerging inorganic thin-film solar cells.

Thermal admittance spectroscopy was used to characterize the defect;[33] however, only shallow defects with ionization energy of about 0.09 eV were measured in both cells (Fig. S22). We turned to quantify the deep defect-induced charge recombination loss in the cell using a modulated electrical transient method (Fig. S23-S24).[34,35] The photovoltage decay rate ($k_r$) of the cells was extracted from the modulated transient photovoltages, as shown in Fig. 4b, which exhibits exponential increases with the voltage. Based on this characteristic, the ideality factor (A) of the control and target cells was fitted to be 1.93 and 1.55, respectively. The significant reduction in the ideality factor indicates that the charge recombination in the CZTSSe absorber has been effectively suppressed. The charge extraction ($\eta_E$) and collection efficiencies ($\eta_C$) correlated with the charge loss in the CZTSSe and buffer/window layer, respectively, were further extracted from the measurements. As shown in Fig. 4c, the $\eta_C$ of these two cells is comparable to each other in the entire voltage regime. In contrast, $\eta_E$ of the target cell has been significantly enhanced and is even higher than the $\eta_C$ in the high-voltage regime. This suggests that the charge non-radiative recombination loss in the CZTSSe film has indeed been reduced. Considering the similar valence band edge and lattice vibration properties between these two samples (Fig. S25 and S26), the improvement in $\eta_E$ is attributed to the reduction in defect concentration, which is estimated to be about one order of magnitude, from $10^{15}$ to $10^{14}$ cm$^{-3}$ (Fig. S27).

We conducted further PL characterization of the CZTSSe films, which emit photons below the bandgap ($E_g$) at low temperatures. Our analysis revealed that the PL red shift (relative to the $E_g$) of the target film is reduced by about 20 meV compared to the control sample (Fig. 4d). This suggests a decrease in the potential fluctuation induced by deep defects.[10,36] To evaluate the surface electric properties, we measured the contacting potential difference (CPD) mapping.[37] As shown in Fig. 4e, both films exhibited similar energy band upward bending behavior in the grain boundary (GB) regions, and their CPD properties were not significantly affected by light illumination (Fig. S28). This implies that both films have negligible surface and GB carrier pinning effect.[38] Consequently, all these results demonstrate that the benefit of the elemental alloying lies in reducing deep defects in the CZTSSe grain interior.

Table 1. Cell performance parameters of our and previously reported record cells.

| Device | | Area (cm$^2$) | PCE (%) | $J_{SC}$ (mA cm$^{-2}$) | $V_{OC}$ (V) | FF | $E_g$ (eV) | $E_g/q$-$V_{OC}$ (V) | $J_{SC}/J_{SC}^{SQ}$ | $V_{OC}$FF/$V_{OC}^{SQ}$FF$^{SQ}$ |
|---|---|---|---|---|---|---|---|---|---|---|
| This work | Control | 0.28 | 11.3 | 35.9 | 0.460 | 0.684 | 1.06 | 0.600 | 0.80 | 0.44 |
|  | Target | 0.28 | 14.6 | 36.6 | 0.550 | 0.724 | 1.08 | 0.530 | 0.82 | 0.55 |
|  | Certified | 0.28 | 14.2 | 36.8 | 0.549 | 0.704 |  | 0.531 | 0.83 | 0.53 |
| NJUPT cell[5] | | 0.11 | 13.0 | 33.6 | 0.529 | 0.729 | 1.11 | 0.581 | 0.76 | 0.51 |
| IBM cell[40] | | 0.42 | 12.6 | 35.2 | 0.513 | 0.698 | 1.13 | 0.617 | 0.81 | 0.46 |
| DGIST cell[41] | | 0.48 | 12.6 | 35.4 | 0.541 | 0.659 | 1.13 | 0.589 | 0.82 | 0.46 |
| IOPCAS cell[42] | | 0.27 | 13.8 | 36.3 | 0.546 | 0.693 | 1.10 | 0.551 | 0.82 | 0.50 |

The current-voltage characteristics of our champion cell are presented in Fig. 4g, showing a total-area PCE of 14.6%. The cell achieved a $J_{SC}$ of 36.6 mA cm$^{-2}$, a $V_{OC}$ of 0.55 V, and an FF of 0.724. Furthermore, the cell obtained a certified PCE of 14.2% in an accredited independent laboratory (National PV Industry Measurement and Testing Center, NPVM, Fig. S29).[39] The main discrepancy between the certified and our lab result, that is the FF, is attributed to the influence of the MgF$_2$ antireflection layer on the electrical probe contacting. The cell's external quantum efficiency (EQE) spectrum (Fig. 4h) yielded an integrated $J_{SC}$ of 38.0 mA cm$^{-2}$ and an

$E_g$ of 1.08 eV. Additional performance parameters of our cell and previously reported cells are summarized in Table 1.[5,40-42] The ratio of our $J_{SC}$ to the Shockley-Queisser limit ($J_{SC}/J_{SC}^{SQ}$) is approximately 0.82, similar to the 12.6% record cells. Moreover, the $V_{OC}FF/V_{OC}^{SQ}FF^{SQ}$ value of our cell reaches 0.55, and the $V_{OC}$ deficit ($E_g/q$-$V_{OC}$) has been reduced to 0.530 V. These two parameters outperform those obtained in our previous 13.8% cell fabricated using high pressure.[42] These results indicate that multinary alloying can more effectively regulate the reaction kinetics and deep defects in CZTSSe, and thus creates more favorable moderate reaction conditions to meet the requirements of large-scale, low-cost industrial manufacturing. Overall, these results represent a significant advancement for CZTSSe solar cells, further promoting the development of emerging inorganic thin-film solar cells (Fig. 4i).[43]

**Acknowledgments:** The authors acknowledge the Excellent Science and Technology Innovation Group of Jiangsu Province in Nanjing University of Science and Technology for their valuable help in the theoretical calculations. This work is supported by the National Natural Science Foundation of China (Nos. 52222212, U2002216, 52227803, 51972332, 52172261) and the Youth Innovation Promotion Association of Chinese Academy of Sciences.

**Author contributions:** J. Shi, J. Wang and Q. Meng conceived the idea. J. Shi did the device simulation, data analysis, device characterization and theoretical calculations and proposed the kinetics mechanism. J. Wang and J. Zhou fabricated the high efficiency solar cells and did the material and device characterization. F. Meng did the STEM characterization and data analysis. X. Xu, K. Yin and L. Lou participated in the device fabrication, optimization and data collection. M. Jiao and B. Zhang participated in the device fabrication. H. Wu, Y. Luo and D. Li participated in the experiment design and discussions. J. Shi and Q. Meng participated in manuscript writing and revising. All authors were involved in the discussions and approved the manuscript.

**Competing interests:** The authors declare that they have no competing interests.